# *A STIFFNESS SENSOR TO HELP IN THE DIAGNOSIS AND THE SURGERY OF ORBITAL PATHOLOGIES*

*V. LUBOZ[1], D. AMBARD[2], F. BOUTAULT[3,2], P. SWIDER[2], Y. PAYAN[1]*

1. TIMC-GMCAO Laboratory, UMR CNRS 5525, Faculté de Médecine, 38706 La Tronche, France
2. Biomechanics Laboratory EA 3697, Purpan University Hospital, 31059 Toulouse, France
3. Maxillofacial Surgery Department, Purpan University Hospital, 31059 Toulouse, France

Simulation has been introduced in computer assisted surgery several years ago to improve planning and surgical techniques. Last improvements in simulation involve more and more biomechanical models. These models can be more or less complex depending on many parameters. Indeed, they can use simple or complex governing equations to simulate the behavior of the object or the phenomena they study. They can also have more or less constitutive elements (either in finite element or in mass spring analysis). In addition, and depending of the clinical applications aimed, patient specific biomechanical model can be required in order to take into account the heterogeneity of soft tissue behavior between each patient.

Modeling soft tissues is particularly difficult and it is still an open problem. Up to now, very few studies have been published in simulating soft tissue behavior. For example, Fung [1] characterized human skin, and Ottensmeyer [5] measured liver properties.

In this paper, the authors present a new sensor used to measure the stiffness of the soft tissues especially in the context of orbital pathologies. This sensor is used (1) to give clinical data on the stiffness of the orbital soft tissues and (2) to improve the results of a finite element (FE) model developed by the authors.

## TECHNICAL ASPECTS

### 1. Clinical background

The main orbital pathology is called exophthalmia. It consists in excessive eyeball protrusion. It is very often related to a thyroid disease called the Basedow illness. The protrusion is mainly due to an increase of the orbital tissue volume and especially of ocular muscles and orbital fat [7]. Exophthalmia can lead to a decrease of visual acuity and sometimes to blindness by elongation of the optic nerve. Clinical observations [7] described the orbital fat tissues during a disthyroidy as the combination of an elastic phase, i.e. the fat fibers (mainly collagen), and a fluid phase, the fat nodules saturated by physiological fluid. An early treatment of the patient endocrinal state may be sufficient but, usually, a surgical reduction of the exophthalmia is needed. A common surgical technique consists in increasing the volume of the orbital cavity, with an ostectomy (*i.e.* a resection) of some parts of the orbital walls [8]. Most of the time, this resection takes place on the maxillary and/or the ethmoid bones. By pushing onto the eye ball, the surgeon evacuates the fat tissues through the holes, towards the sinuses. To avoid critical structures such as the optic nerve, this surgery must be accurate and mini-invasive. Specific tools to improve surgical planning, such as the stiffness sensor, would be helpful.

## 2. Existing work

Besides several clinical studies based on computer tomography (CT) scans and surgical observations, only few recent numerical models of the orbit have been published in the literature. The impact of a foreign body inserted into the eyeball has been studied by Uchio [9] with a FE model of the eyeball. Miller [4] has presented a mass-spring model of the ocular muscles and the eyeball to simulate strabismus surgeries, i.e. with cuts of the orbital muscles. The elastic FE model proposed by Li [2] introduced the first whole orbital soft tissues model and was applied to tumor extraction planning. To our knowledge, [3] was the first work presenting a biomechanical poroelastic model to assist in the planning of the exophthalmia reduction. Poroelastic models are highly dependent upon material properties of the orbital contents such as the Young modulus (for the elasticity), the Poisson ratio (for the compressibility) and the permeability and the porosity of the tissue. Those parameters have been taken into account in the FE model introduced in [3]. To improve the reliability of this model and more generally to offer an approach to improve other predictive models for clinical applications, the authors assume that the measurement of patient related material properties is required. In the context of the exophthalmia surgery and especially for our poroelastic model, the overall stiffness of the orbital contents may be useful to improve the simulation reliability. As a corollary, the investigation of the orbital stiffness could also be clinically relevant for the quantification of the pathology before and during the surgery. To proceed, a specific custom made stiffness sensor has been developed and evaluated on two patients.

# MATERIAL

During the exophthalmia decompression, the surgeon makes a bone resection of the orbit and then pushes on the eyeball to bring it back close to its original position. Consequently, characterizing the stiffness of the orbital tissues seems to be helpful before or during this surgery and for improving our FE model behavior.

The computation of the soft tissues stiffness requires two measurements: the force exerted by the surgeon on the eye ball and the resulting displacement. Consequently, two devices have been coupled to get those data simultaneously: a force sensor and a displacement sensor. Hence, the tissue stiffness can be given in function of the time. The whole stiffness sensor (Fig. 1) has been patented in 2003 [6], and especially the fact that a force sensor and a displacement sensor have been coupled.

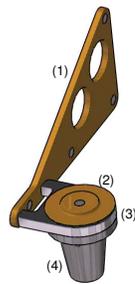

*Fig. 1:*

*Stiffness sensor with (1) rigid body to track the sensor displacement (2) cap on which the surgeon pushes and under which a silicone cylinder is in contact with (3) the membrane measuring the force, and (4) aluminum base.*

The force sensor is made of a general frame in aluminum in contact with the eyeball via a rigid titanium corneal protection. On the opposite extremity, an extensometric gauge, glued underneath a thin membrane (noted (3) on Fig. 1), is affected by the pressure of the surgeon's finger. The force, in Newton, exerted by the surgeon on the sensor is therefore measured via the deformation of this gauge. This force is computed from the measurement of the deformation of the membrane thanks to a prior calibration of the gauge. The force sensor is monitored by a homemade Labview © program.

To measure the displacement, a three-dimensional optical localizer (Polaris ©) is used. Two rigid bodies, thin plastic plates equipped with markers reflecting the infra red beams emitted by the localizer, are needed to locate the sensor. The first rigid body is fixed on the surgical table and is used as the reference. The second rigid body (noted (1) on Fig. 1), fixed on the force sensor manipulated by the surgeon, permits to measure the relative displacement (between this rigid body and the reference one) and to derive the backward displacement of the eyeball. The displacement, in millimeter, of the sensor can be tracked with an accuracy of 0.7mm. The 3D localization is monitored by a homemade Visual C++ © program.

Technically, the sensor must couple those two measuring devices to get the stiffness of the intra orbital tissues. Fig. 1 shows each components of the sensor. The membrane (3) underneath which the gauge is glue is mounted between two aluminum pieces: the fixation device and the base (4). The base is in contact with an ophthalmic shield laid on the eyeball in order to protect it. The silicone cylinder (not visible on the figure) allows spreading the force exerted by the surgeon onto the cap (2). All these pieces can be sterile for a clinical use. The total weight of the sensor is 47 grams (including 10 grams for the rigid body).

The displacement and the force have been recorded with a sample rate of 0.1s on a laptop equipped with a 250MHz processor and 64Mo of memory. As a pilot study, the technique has been used on two patients before and after exophthalmia surgery.

## RESULTS

All the following measurements have been performed by the professor Boutault at the Toulouse Hospital on agreeing patients.

The measurements on the first patient were performed without the optical localizer to evaluate the feasibility of the process in a non localizer equipped operating room (Fig. 2 (a)). The backward displacement $x$ have been visually estimated to 5mm. The measured force $F$ was 20N which provided the first evaluation of the orbital content stiffness: $K = F/x = 4N/mm$.

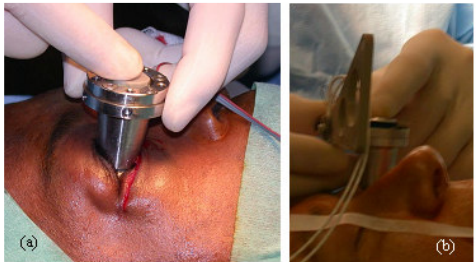

Fig. 2:
Clinical applications of the sensor: (a) without the rigid body and (b) with the displacement measurement device.

In order to be more precise and to get more information about the eyeball displacement, it is necessary to couple the force sensor with the displacement measurements. The complete protocol (force and displacement measurements) has been completed in the second patient (Fig. 2 (b)).

Three loads have been performed for each measurement. Two measurements have been made on the left eye and one on the right eye, before and after surgery. To increase the accuracy of the process, the force sensor was kept in the direction of the axis form by the eyeball and the orbit in order to limit tangential motions.

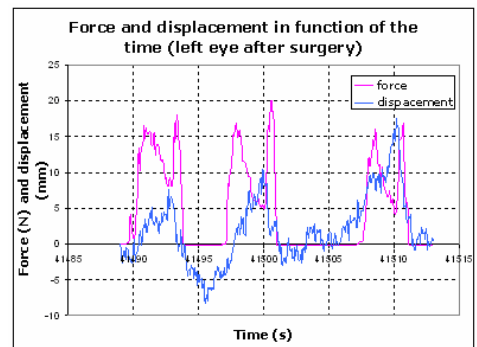

Fig. 3:
Force and displacement measurements for the left eye after exophthalmia decompression.

A typical measurement is plotted by Fig. 3. On this figure, the three loads ant the three displacements resulting can be seen. The force varies from 0N to 19N and the displacement varies from 0 to 17mm. Those two curves have a lot of noise since it is hard to keep a constant force on the sensor and to stay in the same direction while applying this force. Moreover, a negative displacement can be observed in the interval t=41494s and t=41497s. This is due to the fact that the surgeon needed more visibility during the use of the sensor and took the sensor off the eyeball for 3s in order to have a look at his patient. Note that the force is null during this time.

Given that the displacement curve is slightly late compare to the force one, it has been difficult to determine the stiffness curve in function of the time. To get representative values, the authors have assumed that this stiffness was not varying during a measurement since no clinical action was performed at the same time. Indeed, when the sensor is used before the surgery, no hole has been made and the tissue tension remains the same. And when the sensor is used after the surgery, the orbital soft tissues have already been pushed in the sinuses and the stiffness should not vary. Consequently, force values have been chosen (most of them at a curve peak) and the corresponding displacement has been read on the curve. About six representative force/displacement couples have been extracted from the curves resulting of the measurements and the corresponding stiffness have thus been estimated by averaging them. The stiffness values for the 6 measurements made on the second patient are show in Tab. 1.

| Meas. | Orbit | Time | Average stiffness (N/mm) | Stand. dev. (N/mm) |
|---|---|---|---|---|
| 1 | left | pre-op | 2,84 | 1,9 |
| 2 | left | pre-op | 2,18 | 0,4 |
| 3 | right | pre-op | 3,73 | 2,2 |
| 4 | left | post-op | 1,66 | 0,7 |
| 5 | left | post-op | 0,91 | 0,1 |
| 6 | right | post-op | 2,11 | 1,2 |

*Tab. 1:*
*Average stiffness values for the six measurements on the second patient before and after the exophthalmia decompression.*

Recording several displacement/force couples, the average stiffness $K$ was quantified. For the left eye, it was 2.5N/mm before surgery and 1.29N/mm after surgery. The decrease was -24 %. For the right eye, it was 3.73N/mm before surgery and 2.11N/mm after surgery so the decrease was -43%.

## DISCUSSION AND CONCLUSION

The stiffness measurement technique of this study showed satisfying results and it constituted the first objective quantification of the exophthalmia reduction. The measurement technique was relevant to quantify patient-related data. Indeed, the results pointed out a significant difference in the eye-related orbital content and/or the surgery technique (size and location of the osteotomy): -24% for the left eye and -43% for the right eye. According to these values, a difference of approximately 30% can be found between the stiffness before and after the surgery. Moreover, those values allow pointing out the difference of about 80% between the two eyes. This difference may be induced by the fact that the right eye had an exophthalmia more important and that the decompression was larger. To determine if the tendencies showed by those percentages are realistic, the sensor would have to be used on several other patients to get new stiffness values.

On the other hand, and in order to improve the measurements, the noise has to be removed or at least lowered. To reach this aim, improving the control of the displacement of the sensor during the measurement is essential. Before thinking of changing the shape of the sensor, few precautions can be taken. For example, the sensor has to remain in contact most the time during the measurements and the user has to avoid holding it so it can put its whole weight on the eyeball during the entire load.

Though those needed improvements and the fact that the sensor has been tested on only one patient, the values given by the sensor are really interesting. Indeed, they allowed determining an average value of the stiffness of the pathological orbital tissues (around 2.9N/mm) and of the orbital tissues after surgery (around 1.6N/mm). Those latest tissues can also be assumed to be close to non pathological tissues while no measurement has been done for them. By the mean of a large quantitative study on people suffering or not of exophthalmia, those values could be confirmed and may then be useful for diagnostic. Indeed, by determining, with the sensor, if the orbital stiffness for a given patient is too high, it could give a hint of whether or not an exophthalmia decompression has to be done in emergency, later, or not at all.

Besides that, the sensor measurements permitted to partially validate the predictive poroelastic FE model presented in the former study [3]. In this previous work, an average stiffness of 2.85N/mm has been estimated before the decompression thanks to the FE model. The average stiffness value given by the measurements for a pathological orbit was about 2.92N/mm (measures 1 to 3). Consequently, the mean discrepancy between predicted and experimental results was 4%. Though those measures do not come from the same patient, the difference between the two of them is encouraging. Indeed, it seems that our FE model allows a stiffness estimation relatively close to a real measurement.

In conclusion, the results of our pilot study were in agreement with our initial central hypothesis concerning the necessity of measuring patient-related material properties to derive reliable predictive biomechanical model. These promising results will be validated with a larger group of patients in further clinical studies to be usable in clinical settings.
Besides the improvement of the FE model reliability, clinical tests will permit to enlarge the data base in order to develop a robust diagnosis tool: patient-related data, quantification of the degree of pathology, evaluation and improvement of the surgical technique. In parallel, the sensor was also adapted to measure the facial skin stiffness and clinical evaluations are envisaged.

# BIBLIOGRAPHY

[1] FUNG, Y.C.
*Biomechanics: Mechanical Properties of Living Tissues.*
New York: Springer-Verlag. 1993.

[2] LI Z., CHUI C. K., CAI Y., AMRITH S., GOH P. S., ANDERSON J. H., THEO J., LIU C., KUSUMA I., NOWINSKI W. L..
*Modeling of the human orbit from MR Images.*
Proceedings of the MICCAI conference. Springer-Verlag. 2002, 339-347.

[3] LUBOZ V., PEDRONO A., AMBLARD D., SWIDER P., PAYAN Y., BOUTAULT F.
*Prediction of tissue decompression in orbital surgery.*
Clinical Biomechanics, 2004, 19/2 pp. 202-208.

[4] MILLER J. M., DEMER J. L.
*Clinical applications of computer models for strabismus.*
Eds Rosenbaum, Santiago, AP, Clinical Strabismus Management. Pub. W. B. Saunders. 1999.

[5] OTTENSMEYER M.
*In Vivo Data Acquisition Instrument for Solid Organ Mechanical Properties Measurement.*
Proceedings of Medical Image Computing and Computer Assisted Intervention, MICCAI 2001, Utrecht, Netherlands, pp975-982.

[6] PAYAN Y., LUBOZ V., SWIDER P. & AMBARD D.
*Outil et procédé de mesure de la raideur mécanique d'un milieu selon une direction déterminée.*
Université Joseph Fourier de Grenoble / Université Paul Sabatier de Toulouse. Num: 03/51109, 18 Dec. 2003, France.

[7] SARAUX H., BIAIS B., ROSSAZZA
*C. Ophtalmologie, Chap. 22 : Pathologie de l'orbite.*
Ed. Masson., 1987, 341-353.

[8] STANLEY R.J., MCCAFFREY T.V., OFFORD K.P., DE SANTO L.W.
*Superior and transantral orbital decompression procedures. Effects on increased intraorbital pressure and orbital dynamics.*
Arch. Otolaryngol. Head Neck Surg. , 1989, vol. 115, pp. 369-373.

[9] UCHIO E., OHNO S., KUDOH J., AOKI K., KISIELEWICZ L.T.
*Simulation model of an eyeball based on finite element analysis on a supercomputer.*
Journal of Ophtalmology, 1999, 83(10):1101-2.